\documentclass[useAMS,usenatbib,usegraphicx]{mn2e}

\title[Molecular Face-on View of the Galactic Centre]{A Molecular Face-on View of the Galactic Centre Region}
\author[T. Sawada et al.]{Tsuyoshi Sawada,$^{1,2,3}$\thanks{E-mail: sawada@nro.nao.ac.jp}
Tetsuo Hasegawa,$^{2,3}$
Toshihiro Handa$^{3}$ and
R.~J.~Cohen$^{4}$\\
$^{1}$Nobeyama Radio Observatory, 462-2 Nobeyama, Minamimaki, Minamisaku, Nagano 384-1305, Japan\\
$^{2}$National Astronomical Observatory of Japan, 2-21-1 Osawa, Mitaka, Tokyo 181-8588, Japan\\
$^{3}$Institute of Astronomy, University of Tokyo, 2-21-1 Osawa, Mitaka, Tokyo 181-0015, Japan\\
$^{4}$Jodrell Bank Observatory, University of Manchester, Macclesfield, Cheshire SK11 9DL}
\begin{document}

\date{Accepted 2004 January 12. Received 2003 October 11}

\pagerange{\pageref{firstpage}--\pageref{lastpage}} \pubyear{2004}

\maketitle

\label{firstpage}

\begin{abstract}
We present a method to derive positions of molecular clouds along
the lines of sight from a quantitative comparison between $2.6\,{\rm mm}$
CO emission lines and $18\,{\rm cm}$ OH absorption lines, and apply 
it to the central kiloparsecs of the Milky Way.
With some simple but justifiable assumptions, we derive a face-on
distribution of the CO brightness and corresponding radial velocity
in the Galactic centre without any help of kinematical models.
The derived face-on distribution of the gas is elongated and inclined
so that the Galactic-eastern (positive longitude) side is closer to us.
The gas distribution is dominated by a barlike central condensation,
whose apparent size is $500\times 200\,{\rm pc}$.
A ridge feature is seen to stretch from one end of the central condensation,
though its elongated morphology might be artificial.
The velocity field shows clear signs of noncircular motion in the central
condensation.
The `expanding molecular ring' feature corresponds to the
peripheral region surrounding the central condensation with the
Galactic-eastern end being closer to us.
These characteristics agree with a picture in which the kinematics of
the gas in the central kiloparsec of the Galaxy is under a strong
influence of a barred potential.
The face-on distribution of the {\it in situ\/} pressure of the molecular
gas is derived from the CO multiline analysis.
The derived pressure is found to be highest in the central $100\,{\rm pc}$.
In this region, the gas is accumulating and is forming stars.
\end{abstract}


\section{Introduction}

The behavior of molecular gas, in particular its physical conditions and
kinematics, in central regions of galaxies is key information to 
understand the star forming activity which occurs there.
The Galactic centre can be observed in much greater detail compared 
with central regions of other galaxies.
However, its inevitable edge-on perspective sometimes complicates
the interpretation of the data.
In particular, a face-on image of the Galactic centre is 
very hard to construct, though such an image would be very helpful to
understand its kinematics
and to make a comparison with central regions of other galaxies.
Attempts have been made to construct models of gas kinematics
\citep[see, e.g.,][]{liszt1980,binney1991,fux1999}.
Their models give more or less reasonable interpretation of some aspects
of the distribution and motion of the gas, in particular its asymmetric
position-velocity appearance and some outstanding features.
Kinematical models can be used to project position-velocity
diagrams of molecular/atomic line data into a face-on view
\citep[see, e.g.,][]{cohen1983,sofue1995,nakanishi2003}.
This is an indirect method to investigate the spatial distribution of the
gas: it would be invaluable if we could derive positions of
molecular clouds independent of kinematical assumptions.
The molecular content and star forming regions in the Galactic centre
are strongly confined in the central a few hundred parsecs:
the region is called the central molecular zone \citep[CMZ;][]{morris1996}.
Though the kinematics and face-on distribution of the gas within the CMZ
are essential information to study the star forming activity in the
Galactic centre, they are complicated and less well understood than
the outer region (up to the Galactocentric radius of a few kpc)
mainly traced by atomic hydrogen line.

In this paper, we present a method to derive a
molecular face-on view of the Galactic centre without any 
help of kinematical models.
In {\S} \ref{sec-method} we describe the basic methodology.
Using that, we draw a face-on distribution of the molecular gas
from existing data and discuss the resultant face-on view
in {\S} \ref{sec-result}.
Physical conditions of the gas and the validity of parameters 
used in the model are also discussed.

\section{THE METHOD}\label{sec-method}
The essence of our method lies in comparing emission and
absorption spectra toward the Galactic centre.
We compare the CO $2.6\,{\rm mm}$ emission with
the OH $18\,{\rm cm}$ absorption.
Because the Galactic centre region itself is an intense diffuse
$18\,{\rm cm}$ continuum source, strong OH absorption arises 
preferentially from the gas that lies in front of the continuum regions, 
rather than gas that lies behind them.
On the other hand, the CO emission samples the 
gas both in front and back of the continuum sources equally.
Thus the ${\rm OH/CO}$ ratio carries information on the position of
the gas along the line of sight relative to the continuum sources.
In the following, we devise a method to estimate the position
quantitatively by introducing some simple assumptions.

\subsection{Qualitative inspection on molecular gas distribution}
\label{subsec-dataproc}
We used the Columbia-U.\ Chile ${}^{12}{\rm CO}\; J=1-0$
data \citep{bitran1997} and a large-scale absorption survey of
the OH main lines ($1665\, {\rm MHz}$ and $1667\, {\rm MHz}$)
made by \citet{boyce1994}.
Both data were binned over successive $5\,{\rm km\, s^{-1}}$
velocity ranges.
The CO data, for which the 
original resolution and grid spacing are respectively
9 arcmin and 7.5 arcmin, were smoothed to 10-arcmin resolution
and resampled with a 12-arcmin grid to match the OH data.
Hereafter we mainly used the $1667\, {\rm MHz}$ line
for the OH absorption.
However, because of the large velocity spread of
the molecular clouds in the Galactic centre, the 
positive-velocity end of the $1667\, {\rm MHz}$ line
blends together with the negative-velocity end of the
$1665\, {\rm MHz}$ line.
Thus we replaced the $v_{\rm LSR} > 100\,{\rm km\, s^{-1}}$
part of the $1667\,{\rm MHz}$ line with the $1665\, {\rm MHz}$ data
scaled by 1.37, which is the mean ratio of the absorption
depth between the two lines in the velocity range 
$25 \le v_{\rm LSR} \le 100\,{\rm km\, s^{-1}}$.

\begin{figure}
\includegraphics[width=84mm]{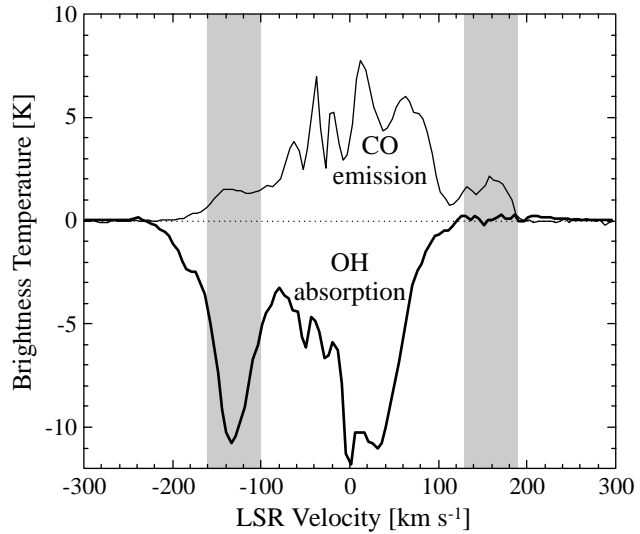}
\caption{Sample spectra of CO emission ({\it thin line\/}) and
OH absorption ({\it thick line\/}) at $(\ell,b) = (-0\fdg 2,0\fdg 0)$.
Shadowed velocity components demonstrate
the difference of cloud position along the line of sight (see text).}
\label{fig-prfcmp}
\end{figure}
Figure \ref{fig-prfcmp} shows the processed spectra of the OH
absorption and the CO emission at $(\ell,b)=(-0\fdg 2,0\fdg 0)$.
This figure demonstrates that the ${\rm OH}/{\rm CO}$ ratio
varies significantly as a function of the radial velocity, $v_{\rm LSR}$.
For example, we may draw attention to the velocity components near
$v_{\rm LSR}\simeq -130\,{\rm km\, s^{-1}}$ and
$v_{\rm LSR}\simeq 160\,{\rm km\, s^{-1}}$
(shadowed in Fig.\ \ref{fig-prfcmp}),
both of which belong to the so-called `expanding molecular ring'
\citep*[EMR;][]{kaifu1972,scoville1972}.
The CO intensities at these components are almost the same.
This suggests that the amounts of molecular gas are similar
in each component.
On the other hand, the OH absorption depths
are strikingly different.
The negative-velocity component shows deep absorption,
while there is no distinct absorption around the
positive-velocity component.
We immediately deduce from this fact that
the negative-velocity component is located
in front of strong continuum source surrounding the Galactic centre,
while the positive-velocity component is behind it.
This logic led \citet{kaifu1972} to conclude that this feature
is expanding away from the Galactic centre.

\begin{figure}
\includegraphics[width=84mm]{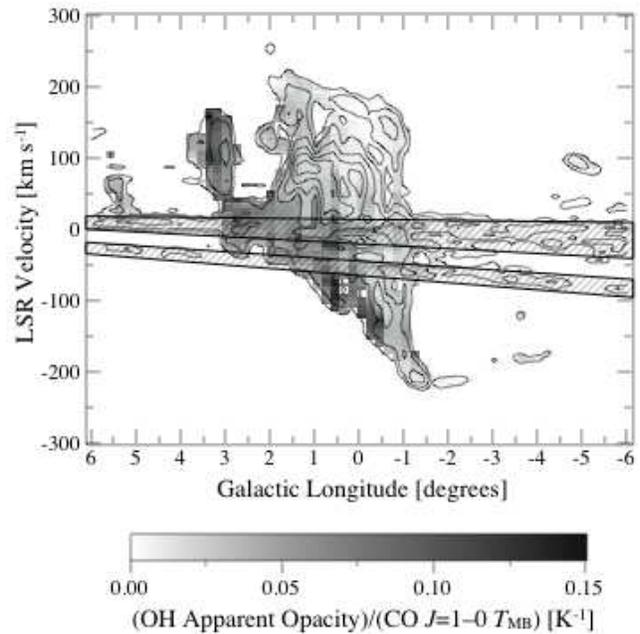}
\caption{The ratio between the OH apparent opacity
and the ${\rm CO}\; J=1-0$ intensity at $b=0\fdg 0$.
Contours are $T_{\rm MB}({\rm CO}) = 1, 2, 4, 7, 10,$
and $14\, [{\rm K}]$.
The ratio is shown in the region where $T_{\rm MB}({\rm CO})$
exceeds $1\,[{\rm K}]$.
Hatched velocity ranges were excluded from the analysis because of
contamination by clouds well outside the Galactic centre region.}
\label{fig-ohcoratio}
\end{figure}
Figure \ref{fig-ohcoratio} shows the longitude-velocity
($\ell$-$v$) diagram of the ratio between the OH apparent opacity
($\tau_{\rm app} \equiv T_{\rm abs}/T_{\rm cont}$;
where $T_{\rm abs}$ and $T_{\rm cont}$ are
line absorption and continuum antenna temperatures, respectively) and 
the ${\rm CO}\; J=1-0$ line intensity.
The ratio has a clear trend; it is
smaller in $\ell < 0\degr $ and $v_{\rm LSR} > 0\,{\rm km\, s^{-1}}$,
while it is larger in
$\ell > 0\degr $ and $v_{\rm LSR} < 0\,{\rm km\, s^{-1}}$
and in the feature with extremely large velocity width at
$\ell \simeq 3\degr$, which is called Bania's Clump 2 \citep{bania1977}.
The ratio in the velocity range
$-60\,{\rm km\, s^{-1}} \la
v_{\rm LSR} \la 10\,{\rm km\, s^{-1}}$
is affected by the foreground gas in the Galactic disc and does not show
the intrinsic value of the Galactic centre.
Thus the hatched velocity ranges in Fig.\ \ref{fig-ohcoratio}
are excluded from the following analysis.

\subsection{Deriving distances to clouds}\label{sec-derivedist}
Based upon the logic shown in the previous section,
we extend it to quantitative estimation of molecular gas distribution.
In order to determine the distances to molecular clouds quantitatively,
we adopt the following four assumptions:
\\
(1) At a given $\ell$, emission observed at each velocity bin comes
from a single position along the line of sight.\\
(2) The CO line intensity $T_{\rm CO}$ at a given velocity 
is proportional to the amount of molecular gas in unit velocity width;
and the OH opacity $\tau_{\rm OH}$ (not the `apparent' 
one but the real one) at a given velocity is also proportional to the 
amount of molecular gas in unit velocity width.
Consequently, $\tau_{\rm OH}= Z T_{\rm CO}$ where $Z$ is a constant.\\
(3) The excitation temperature of OH
[$T_{\rm ex} ({\rm OH})$] is uniform.\\
(4) The $18\,{\rm cm}$ continuum emission is optically thin
and arises from a distributed volume emissivity, $j(r)$
[$r$ is the Galacto\-centric radius],
that is axisymmetric about the Galactic centre as modelled in
{\S} \ref{subsec-contem}.

Now $Z$ and $T_{\rm ex}({\rm OH})$ are unknown parameters.
How to determine them is described in {\S} \ref{subsec-paramchoise};
validity of them and the assumption (1) are discussed in
{\S} \ref{subsec-paramvalid}.

When a cloud whose OH opacity is $\tau_{\rm OH}$ is located
at $s = s_0$ ($s$ is the position along the line of sight),
the cloud absorbs the continuum intensity behind it,
i.e., the continuum emissivity integrated along the line of sight
behind the cloud, $\int_{-\infty}^{s_0} j(r) {\rm d}s$
using the assumption of optically thin continuum emission.
The absorption depth is written as
$f [1-\exp(-\tau_{\rm OH})]
[\int_{-\infty}^{s_0}j(r) {\rm d}s-T_{\rm ex}({\rm OH})]$
where $f$ is the beam filling factor of OH absorbing gas.
The total continuum intensity observed by us is
$\int_{-\infty}^{s_\odot}j(r) {\rm d}s$, and the apparent opacity
$\tau_{\rm app}$ of the cloud is expressed as
\begin{equation}
\label{eq-tauapp}
\tau_{\rm app} =
\frac{f \left[1-\exp(-\tau_{\rm OH})\right]
\left[\int_{-\infty}^{s_0}j(r) {\rm d}s-T_{\rm ex}({\rm OH})\right]}
{\int_{-\infty}^{s_\odot}j(r) {\rm d}s}.
\end{equation}
Since $\tau_{\rm app}$ and $T_{\rm CO}$ (and thus
$\tau_{\rm OH} = ZT_{\rm CO}$) are
known through the observations, we can obtain the value of
$\int_{-\infty}^{s_0} j(r) {\rm d}s$ if we know $f$, $Z$, and
$T_{\rm ex}({\rm OH})$.
Then $s_0$ is derived from the value of $\int_{-\infty}^{s_0} j(r) {\rm d}s$
using a distribution model of $j(r)$.

\subsection{Modelling the distribution of continuum emissivity}\label{subsec-contem}
\begin{figure}
\includegraphics[width=84mm]{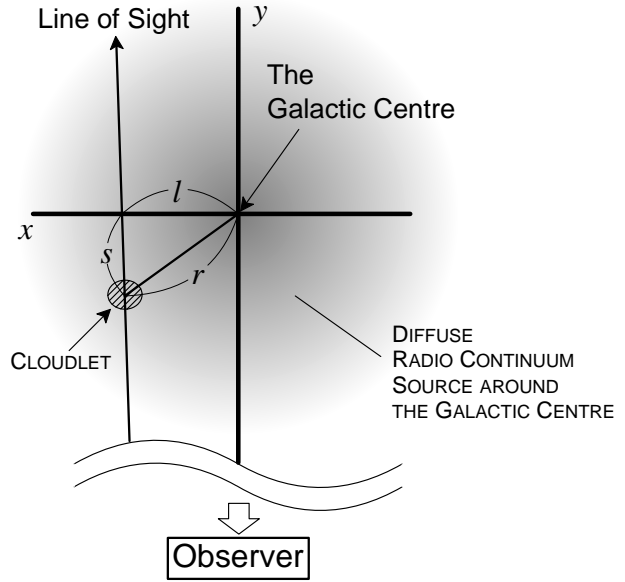}
\caption{The schematic relation of geometrical parameters in the face-on
view of the Galactic centre region.}
\label{fig-punch}
\end{figure}
\begin{figure}
\includegraphics[width=84mm]{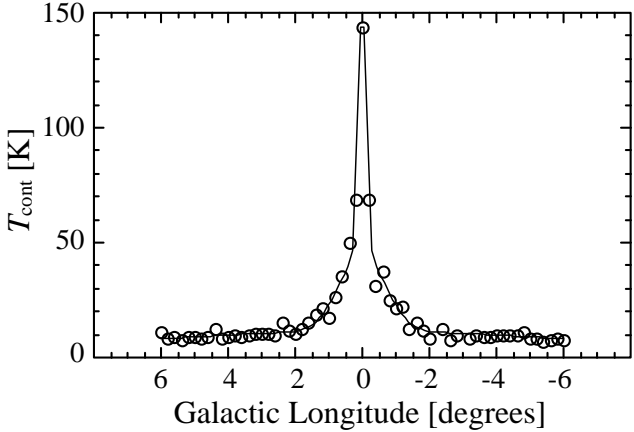}
\caption{Longitudinal distribution of the $18\,{\rm cm}$ continuum
in units of antenna temperature
at $b=0\fdg 0$ by \citet{boyce1994} ({\it open circles\/}) and
a fit by 3 Gaussians ({\it solid line\/}).
The fitting function is $T_{\rm cont} = \sum_{i=1}^{3}
a_i\exp\left(-{\ell^2}/{(2{\sigma_i}^2)}\right)$
where
$a_1=98.3\,[{\rm K}]$,
$a_2=34.0\,[{\rm K}]$,
$a_3=11.4\,[{\rm K}]$;
$\sigma_1=0\fdg 120$,
$\sigma_2=0\fdg 677$,
$\sigma_3=7\fdg 18$.}
\label{fig-contfit}
\end{figure}

We assume that the $18\,{\rm cm}$ continuum emissivity $j(r)$
around the Galactic centre
is described as a sum of several axisymmetric Gaussians:
\begin{eqnarray}
j(r) &=& \sum_{i=1}^{N} \frac{a_i}{\sqrt{2\pi}\sigma_i}
 \exp\left(-\frac{r^2}{2{\sigma_i}^2}\right)\\
 &\simeq& \sum_{i=1}^{N} \frac{a_i}{\sqrt{2\pi}\sigma_i}
 \exp\left(-\frac{\ell^2+s^2}{2{\sigma_i}^2}\right).
\end{eqnarray}
Figure \ref{fig-punch} shows the schematic relation of geometrical
parameters.
The Galactic longitude $\ell$ is in degrees.
Here $r$, $s$, and $\sigma_i$ are in units of projected distance
corresponding to $1\degr$ at the distance to
the Galactic centre: i.e., $150\,{\rm pc}$ at $8.5\,{\rm kpc}$.
The observed longitudinal distribution of the continuum brightness
$T_{\rm cont}(\ell)$ can be written as
\begin{equation}
T_{\rm cont}(\ell) = \int_{-\infty}^{s_\odot} j(r)\, {\rm d}s.
\end{equation}
Since the continuum emissivity beyond the solar circle should
be negligible compared with that in the centre,
$\int_{-\infty}^{s_\odot} j(r)\, {\rm d}s$ can be replaced with
$\int_{-\infty}^{\infty} j(r)\, {\rm d}s$.
Thus
\begin{eqnarray}
T_{\rm cont}(\ell) &\simeq& \int_{-\infty}^{\infty}
\left[ \sum_{i=1}^{N} \frac{a_i}{\sqrt{2\pi}\sigma_i}
 \exp\left(-\frac{\ell^2+s^2}{2{\sigma_i}^2}\right) \right] {\rm d}s\\
&=& \sum_{i=1}^{N}
a_i\exp\left(-\frac{\ell^2}{2{\sigma_i}^2}\right).\label{eq-Icont}
\end{eqnarray}
Now $a_i$ and $\sigma_i$ can be drawn so that
the Eq.\ (\ref{eq-Icont}) reproduces the observed
longitudinal distribution of the continuum brightness.
It is found that observed continuum distribution is well fitted
by up to three components ($N=3$).
For $b = 0\fdg 0$, we obtain
$a_1=98.3$, $a_2=34.0$, $a_3=11.4$;
$\sigma_1=0.120$, $\sigma_2=0.677$, $\sigma_3=7.18$.
Parameters $a_i$ are in Kelvins, and $\sigma_i$ are in degrees.
Figure \ref{fig-contfit} shows the result of our fit.
The model reproduces the observed data quite well.

Since $\sigma_1$, $\sigma_2$ are small enough
($\la 100\,{\rm pc}$) and $a_1$, $a_2$ are large,
the continuum distribution due to the first ($i=1$) and
the second ($i=2$) components is well defined.
In some galaxies seen in nearly face-on perspective, such as
M83 \citep{ondrechen1985} and
IC 342 \citep*{crosthwaite2000},
$\lambda \simeq 20\,{\rm cm}$ continuum emission in
the central kiloparsec consists of strong central source
and extended emission seen in resolutions of hundreds of
parsecs.
For such continuum distribution, axisymmetric emissivity
distribution is a good approximation in the first order.
On the other hand, there are galaxies whose central
$\lambda \simeq 20\,{\rm cm}$ emission is
dominated by ring structure,
such as NGC 1097 \citep*{hummel1987} and NGC 1365 \citep*{sandqvist1995}.
This is not the case for the Milky Way Galaxy
because the $18\,{\rm cm}$ continuum brightness
in the Galactic centre, our edge-on view,  has an intense central peak
with a monotonic decrease away from the centre (see Fig.\ \ref{fig-contfit}).
The fact that the central peak, which should originate in the Sgr A
region, is seen through the foreground medium supports our assumption
that the $18\,{\rm cm}$ continuum is optically thin.

On the other hand, because of large $\sigma_3$ and small $a_3$,
the third ($i=3$) component should suffer from possible
non-axisymmetric distribution on the largest scale and/or
individual sources; furthermore, small ${\rm d}T_{\rm cont}/{\rm d}s$ causes
large errors in the derived positions of clouds.
Therefore, in longitudes where the contribution from
the first and the second components is negligible
(i.e., $| \ell | \ga 1\fdg 5$),
the positions of clouds obtained using this model
are rather uncertain.

\subsection{Choice of the $Z$ and $T_{\rm ex}({\rm OH})$ values}\label{subsec-paramchoise}
\begin{figure*}
\includegraphics[width=126mm]{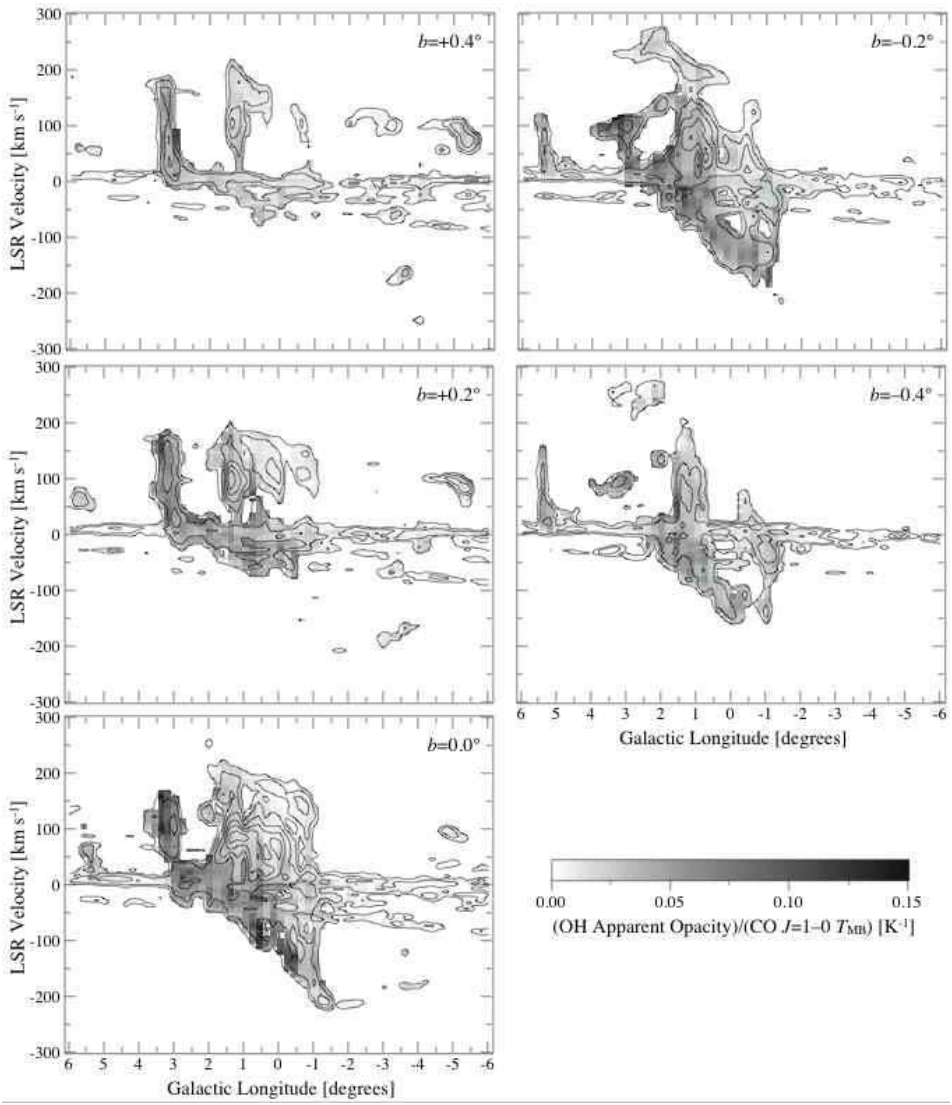}
\caption{Longitude-velocity distribution of the ${\rm OH}/{\rm CO}$ ratio,
shown in the same way as Fig.\ \ref{fig-ohcoratio}.}
\label{fig-lvall}
\end{figure*}

\begin{figure*}
\includegraphics[width=126mm]{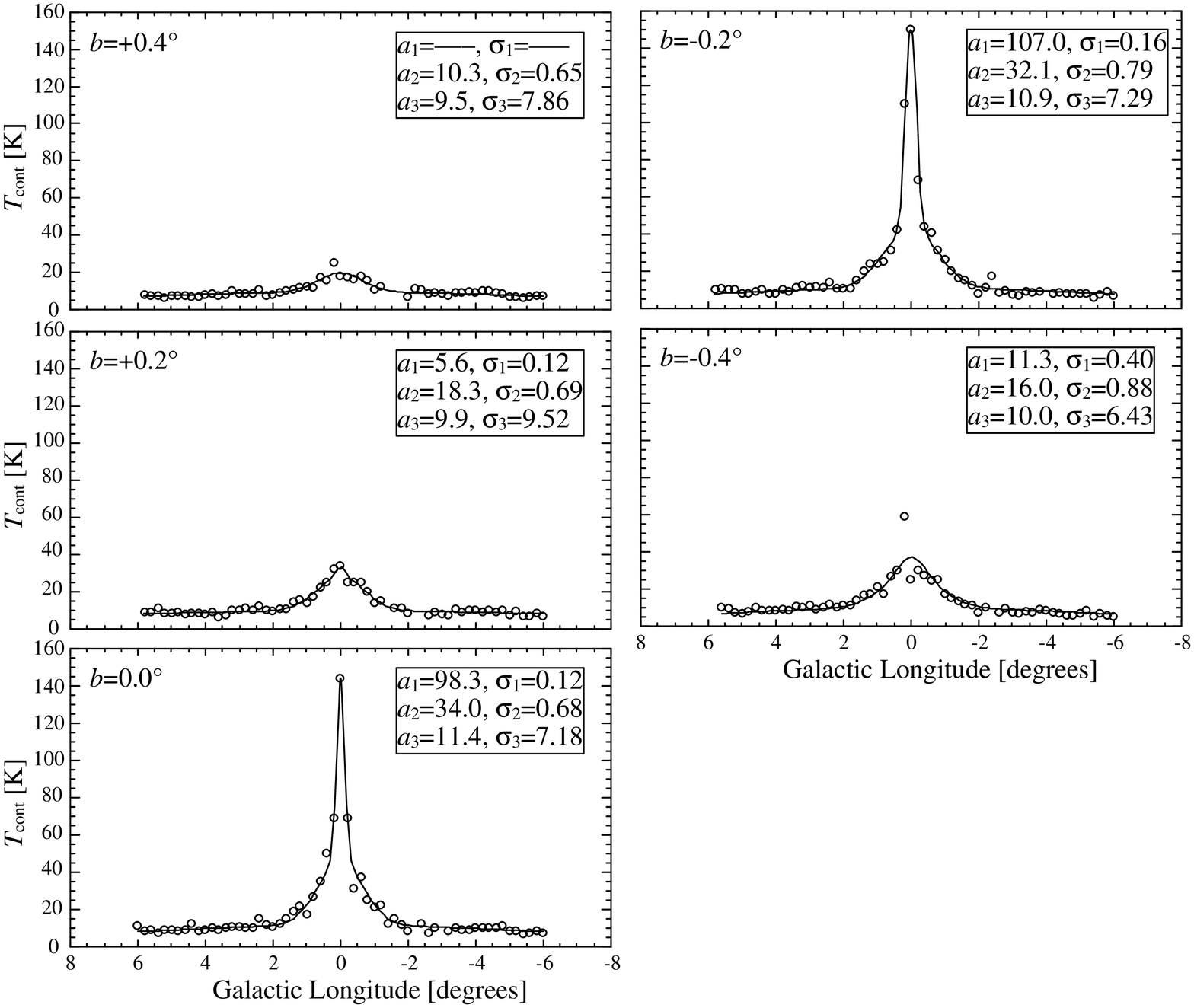}
\caption{Longitudinal distribution of the $18\, {\rm cm}$ continuum
brightness and model fits
shown in the same way as Fig.\ \ref{fig-contfit}.
The fit parameters $a_i$ and $\sigma_i$ ($i=1,2,3$) are in
Kelvins and degrees, respectively.}
\label{fig-contall}
\end{figure*}

Following the procedure, we can draw a face-on ($x$-$y$)
distribution of the molecular gas by putting each data point of the
$\ell$-$v$ diagram (Fig.\ \ref{fig-ohcoratio}) on to the $x$-$y$
plane with a projection of $(\ell,s) \to (x,y)$.
However, $f$, $Z$, and $T_{\rm ex}({\rm OH})$ are still unknown.
Here we assume $f=1$: the validity of this assumption is discussed in
{\S} \ref{subsubsec-validf}.
There is no bottom-up scheme to determine $Z$ and $T_{\rm ex}({\rm OH})$.
To determine $Z$ and $T_{\rm ex}({\rm OH})$,
we employ a trial-and-error scheme, making face-on maps at
$b=+0\fdg 4, +0\fdg 2, 0\fdg 0, -0\fdg 2,$
and $-0\fdg 4$ with various values of $Z$ and $T_{\rm ex}({\rm OH})$.
The data used, i.e., the ${\rm OH}/{\rm CO}$ $\ell$-$v$ diagrams and
the continuum distribution, are shown in Figures \ref{fig-lvall} and
\ref{fig-contall}.
Trials have been done for $Z=0.04$ to $0.70\,[{\rm K}^{-1}]$ and
$T_{\rm ex}({\rm OH}) = 0$ to $10\,[{\rm K}]$
\setcounter{footnote}{0}\footnote{
$T_{\rm ex}({\rm OH})$ is measured as an excess over the
cosmic microwave background (CMB).}.
%

We have chosen an appropriate set of $(Z,T_{\rm ex}({\rm OH}))$
so that the following three conditions are satisfied:\\
(1) The resultant face-on distribution of the CO brightness is not too
asymmetric between the near and far sides with respect to the centre,\\
(2) The features extending above and below the Galactic plane are placed
in similar face-on positions at different latitudes, and\\
(3) Most of the CO emission has a solution for the position $s_0$.
\begin{figure*}
\includegraphics[width=147mm]{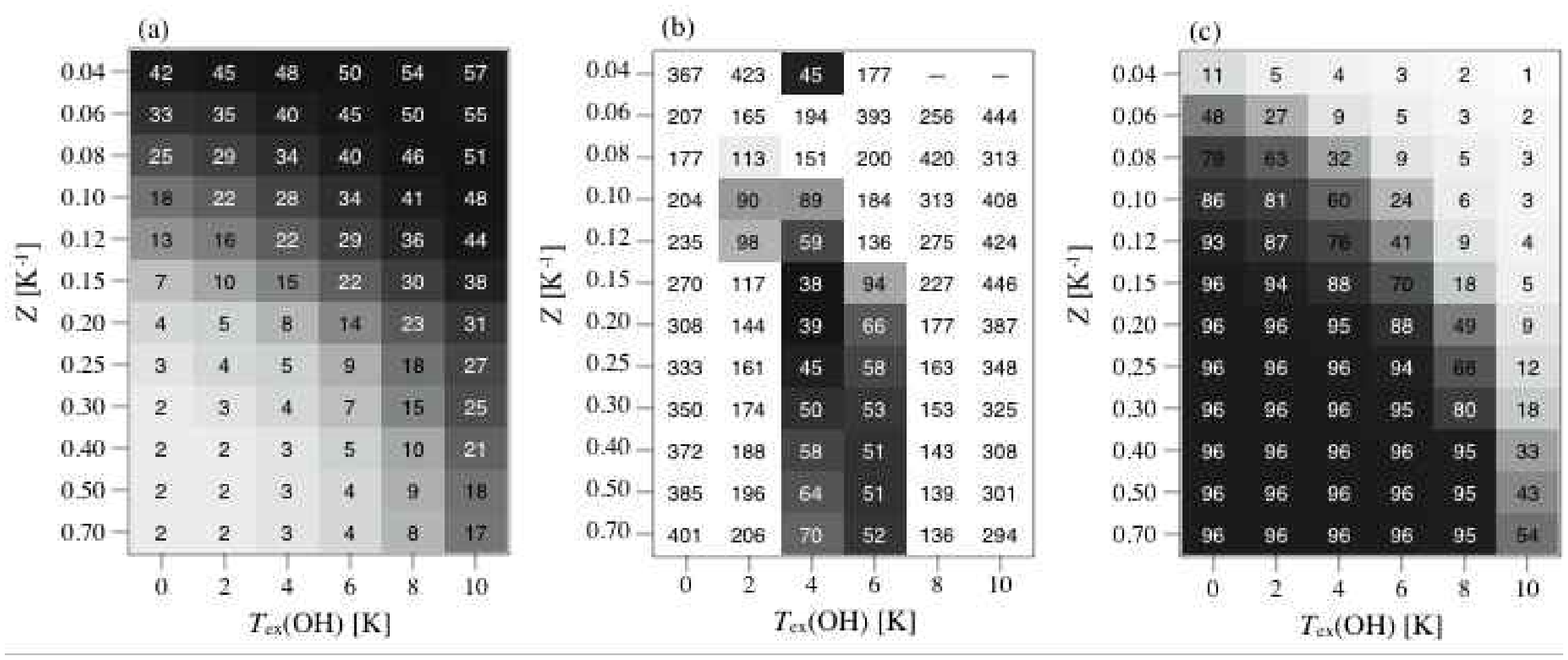}
\caption{(a) Percentage of the CO emission within $|\ell| \le 1\degr$
which is located in the near side.
The unit is per cent.
(b) The standard deviation of the centres of CO emission in
a cloud complex ($1\fdg 0 \le \ell \le 1\fdg 6$,
$40 \le v_{\rm LSR} \le 130\,{\rm km\, s^{-1}}$)
at $b=0\fdg 0, \pm 0\fdg 2, \pm 0\fdg 4$ in parsecs.
(c) Percentage of the CO emission in Clump 2
($2\fdg 6 \le \ell \le 3\fdg 4$, $0\fdg 0 \le b \le 0\fdg 4$,
$30 \le v_{\rm LSR} \le 200\,{\rm km\, s^{-1}}$)
which can be placed within the inner $3750\,{\rm pc}$.
The unit is per cent.
Desirable parameter space is shadowed.}
\label{fig-params}
\end{figure*}
%

In order to check the condition (1), the CO emission within
$|\ell| \le 1\degr$ is treated.
The percentage of the CO emission located in the near side ($s_0>0$) is
displayed in Figure \ref{fig-params}a.
The central molecular emission has a well-known asymmetry
with respect to $\ell = 0\degr$:
\citet{bally1988} mentioned that about three-fourths of molecular
($^{13}{\rm CO}$ and CS) emission comes from positive longitudes.
Therefore asymmetry to a similar degree along the line of sight is
acceptable.
Sets of large $Z$ and small $T_{\rm ex}({\rm OH})$ are rejected
since they would put most of the emission on the far side.

A cloud complex around
$(\ell,v_{\rm LSR})\simeq (1\fdg 3, +100\,{\rm km\,s^{-1}})$
[hereafter the `$1\fdg 3$ region'] is chosen to check the condition (2).
At first we have derived the centres of emission along the lines of sight
in $b = 0\fdg 0, \pm 0\fdg 2, \pm 0\fdg 4$.
Then the standard deviation of the centres is calculated:
Figure \ref{fig-params}b shows the results.
The standard deviation becomes small when we adopt
$Z\ga 0.10\,[{\rm K^{-1}}]$
and $T_{\rm ex}({\rm OH}) \simeq 4\,[{\rm K}]$.
The gas at high latitudes goes farther with respect to that at
$b=0\fdg 0$ if we adopted smaller values for $T_{\rm ex}({\rm OH})$;
the reverse is also true.

For the condition (3), we deal with Clump 2.
Because of its large ${\rm OH}/{\rm CO}$ ratio, Clump 2 is apt to have
no solution; i.e., the large absorption depth cannot be reproduced even if
the cloud is located just in front of us.
Figure \ref{fig-params}c shows the percentage of the CO emission which
can be placed within
$-3750 < s_0 < 3750\,{\rm pc}$ for the used sets of
$(Z,T_{\rm ex}({\rm OH}))$.
If we use small $Z$ or large $T_{\rm ex}({\rm OH})$,
significant emission is lost.

By combining these conditions, we have chosen
$Z=0.15\pm 0.03\,{\rm [K^{-1}]}$ and 
$T_{\rm ex}({\rm OH})=4\pm 1\,{\rm [K]}$.
As noted in {\S} \ref{subsec-contem}, the derived distances to the clouds
are uncertain in $|\ell|\ga 1\fdg 5$.
Thus the constraint on the parameters given by condition (3) using
Clump 2 ($\ell \simeq 3\degr$) is weaker than those given by 
conditions (1) and (2).
However, parameter space allowed by condition (2) is involved in
that given by condition (3).
Thus  we can obtain
$(Z,T_{\rm ex}({\rm OH}))=(0.15\pm 0.03\,{\rm [K^{-1}]},4\pm 1\,{\rm [K]})$
using the conditions (1) and (2) alone; condition (3) is then
automatically satisfied.

\section{RESULTS AND DISCUSSION}\label{sec-result}
\subsection{Distribution and kinematics of the gas}\label{subsec-result}
\subsubsection{Overall structure}
\begin{figure*}
\includegraphics[width=126mm]{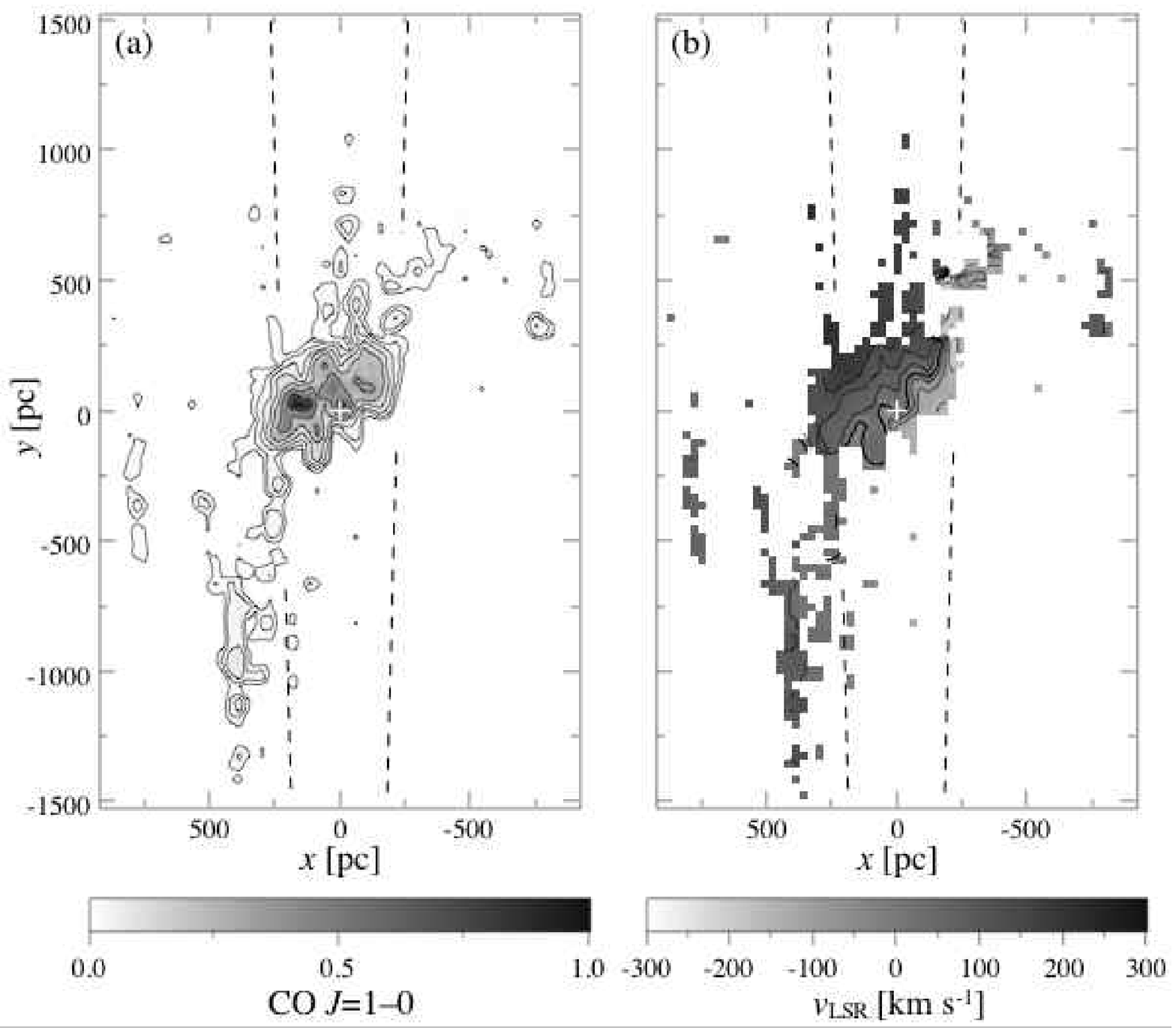}
\caption{The resultant molecular face-on view of the
Galactic centre at $b = 0\fdg 0$.
The parameter set
$(Z,T_{\rm ex}({\rm OH})) = (0.15\,{\rm K^{-1}},4\,{\rm K})$
is used.
Distribution of ${\rm CO}\; J=1-0$ emission (normalized by the peak value)
is shown in ({\it a\/}).
Contours are 0.01, 0.02, 0.04, 0.08, 0.16, 0.32, and 0.64
of the peak value.
Corresponding radial velocity is shown in ({\it b\/}).
Distribution of the velocity is shown in the region where the CO emission
is more intense than 1 per cent of the peak.
Contours are $-200$ to $+200\,{\rm km\, s^{-1}}$
with a spacing of $50\,{\rm km\, s^{-1}}$.
Thick contour is $v_{\rm LSR}=0\,[{\rm km\, s^{-1}}]$.
The solar system is located at $(x,y)=(0,-8500)$.
White crosses in both panels show the centre, $(x,y)=(0,0)$.
Dashed lines indicate $\ell = \pm 1\fdg 5$.}
\label{fig-result}
\end{figure*}
Figure \ref{fig-result} shows the resultant molecular face-on
view of the Galactic centre region at $b = 0\fdg 0$ seen from
the direction of the north Galactic pole.
The CO brightness of each pixel in Fig.\ \ref{fig-ohcoratio} is projected
on to the $x$-$y$ plane and smoothed by
$\sigma=20\,{\rm pc}$ Gaussian.
Figures \ref{fig-result}{\it a\/} and \ref{fig-result}{\it b\/} show the
distributions of CO brightness and corresponding radial velocity,
respectively.
Dashed lines in the Figure indicate $\ell = \pm 1\fdg 5$:
as mentioned in {\S} \ref{subsec-contem},
the obtained face-on maps are less reliable outside them.

The resultant CO distribution mainly consists of a central condensation
(corresponding to the CMZ)
and a thin ridge feature at $(x,y)\simeq(400\,{\rm pc},-1000\,{\rm pc})$
that stretches almost along the line of sight from the Galactic-eastern
(positive $x$ in Fig.\ \ref{fig-result}) end of the
central condensation.
As seen in Fig.\ \ref{fig-ohcoratio},
high ${\rm OH}/{\rm CO}$ ratio is more widespread in
the positive longitudes.
This comes about because the face-on CO distribution
is inclined so that the gas in the positive $x$
(corresponding to positive $\ell$) lies closer to us,
as qualitatively suggested by \citet{cohen1976}.
It has long been argued that the Milky Way is a barred galaxy.
Recent studies agree with a picture in which the Galactic-eastern
(positive longitude) end of the bar is closer to us
\citep[see, e.g.,][and references therein]{morris1996,gerhard1999}.
Our face-on map shows, as a whole, the same trend as this picture.

Some barred galaxies, such as NGC 1097
\citep{ondrechen1983,hummel1987} and
M83 \citep{ondrechen1985},
show $\lambda \simeq 20\,{\rm cm}$ continuum enhancements
in their bars; in particular, in the dust lanes.
If the Milky Way is barred and the $18\,{\rm cm}$ continuum emissivity
is elongated along its bar like these galaxies,
then the actual distribution of molecular gas would be 
{\it more\/} inclined than
that obtained here.
Therefore the trend of inclination is very certain.

\subsubsection{Central condensation}
The central condensation dominates the CO emission in the Galactic centre
region.
It is elongated: its apparent size is approximately
$500\times 200\,{\rm pc}$,
and should be compared with `twin peaks'
in central regions of barred galaxies \citep{kenney1992}.
The minor axis length of the condensation might be smaller since
the face-on map involves positional errors along the lines of sight
as noted later in {\S} \ref{subsubsec-clump2}.
The major axis of the condensation is inclined with respect to the line of
sight ($x=0$) by $\simeq 70\degr$ so that the Galactic-eastern side is
closer to us.
This angle does not change significantly in the acceptable parameter space
of $(Z,T_{\rm ex}({\rm OH}))$.
The condensation includes Sgr A ($l\simeq 0\fdg 0$),
Sgr B ($\ell \simeq 0\fdg 6$), and Sgr C ($\ell \simeq -0\fdg 5$) molecular
cloud complexes and the $1\fdg 3$ region.
The resultant face-on distribution of radial velocity
(Fig.\ \ref{fig-result}{\it b\/}) clearly shows that the gas motion in the
condensation is strongly noncircular.
The gas in the far-side has larger receding velocity,
while the gas in the near-side is rather approaching.
This velocity structure can be explained if the gas orbit is elongated
along the major axis of the central condensation.
It is a clear sign of a bar:
a similar trend is often seen in both
numerical simulations of gas kinematics in a barred potential
and observations of barred galaxies
\citep*[see, e.g.,][]{athanassoula1992,lindblad1996}.
This fact supports the arguments that the Milky Way is barred.

\subsubsection{Bania's clump 2}{\label{subsubsec-clump2}
The ridge, which  lies at $(x,y)\simeq(400\,{\rm pc},-1000\,{\rm pc})$,
corresponds to the feature called Bania's Clump 2.
Clump 2 has long been noticed because of its peculiar characteristics:
widespread velocity structure and large latitudinal extent.
\citet{stark1986} considered that
Clump 2 can be understood as an end-on projection of
a dust lane or an inner spiral arm.
\citet{fux1999} argued that, however, a dust lane which associates with
the Galactic-eastern side of the bar
does not correspond to the Clump 2, but to the
`connecting arm' feature \citep{rougoor1960}
based on numerical simulation.
\citet{fux1999} discussed Clump 2 and 
interpreted it as gas clouds which are crossing a dust lane shock.

In our face-on map, the ridge corresponding to Clump 2
resembles offset ridges (or dust lanes) often seen in
barred spiral galaxies.
Naively speaking, the entire structure,
a central condensation and a possible offset ridge-like feature,
is similar to those seen in central regions
of barred galaxies \citep[see, e.g.,][]{kenney1992}.
It also mimics numerically simulated distributions of materials
orbiting in a barred potential \citep[see, e.g.,][]{athanassoula1992}.

However, the elongation of the ridge along the line of sight might be
artificial:
the face-on maps obtained here inevitably involve positional errors
along the lines of sight.
The nominal error reduces around the centre
since the gradient of continuum brightness (${\rm d}T_{\rm cont}/{\rm d}s$,
namely, continuum emissivity $j$)
becomes steeper; and vice versa.
The
$1\sigma$ deviations in $T_{\rm CO}$ and $\tau_{\rm app}$
typically cause $\simeq 200\,{\rm pc}$ error around Clump 2,
while the error in the central condensation is a few tens of parsecs.
A localized source of radio continuum emission embedded in the
molecular cloud could also lead, in our model, to an apparently more 
widespread distribution
of the cloud in the derived face-on map than the actual one.

Therefore our results cannot determine which of the interpretations
[\citet{stark1986} and \citet{fux1999}] is relevant.
Nevertheless, it is certain that Clump 2 lies in front of
the centre, since the ${\rm OH}/{\rm CO}$ ratio there is
remarkably high.

\subsubsection{`Expanding Molecular Ring' feature}
The so-called EMR feature was considered as an expanding (and rotating) ring
in the early works \citep{kaifu1972,scoville1972},
as obvious from its name, based on its tilted oval $\ell$-$v$ appearance.
Later, another interpretation that the apparent expansion
results from elongated orbit of the gas was proposed.
\citet{binney1991} claimed that the $\ell$-$v$ locus of the EMR,
a parallelogram rather than an ellipse, can be
understood as a projection of the so-called $x_1$ orbit
\citep{contopoulos1977} in a barred potential.
As noted in the present (see {\S} \ref{subsec-dataproc}) and past
\citep[see, e.g.,][]{kaifu1972} works,
the negative-velocity component of the EMR
lies in front of the centre, while the positive-velocity component
is located behind of the centre.
Both interpretations, an expanding ring and an orbit in a barred potential,
were constructed to match this geometry.

\begin{figure}
\includegraphics[width=84mm]{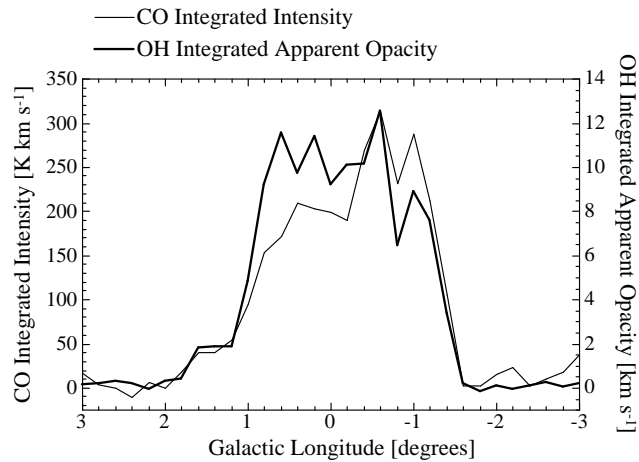}
\caption{Longitudinal distribution of ${\rm CO}\; J=1-0$ integrated
intensity ({\it thin line\/}) and OH integrated apparent opacity
({\it thick line\/}) for negative-velocity component of the EMR.
The velocity range for the integration is
$-300 \le v_{\rm LSR}\,[{\rm km\, s^{-1}}] \le%
-65 + 5 \ell$ ($\ell$ is in degrees):
approaching toward us faster than the hatched velocity ranges
in Fig.\ \ref{fig-ohcoratio}.}
\label{fig-emr}
\end{figure}
In the face-on map obtained in this work, the EMR (in particular
negative-velocity component) is not clearly separated from the central
condensation.
We investigate the origin of the EMR, an energetic expansion or an orbital
crowding, based on the ${\rm OH}/{\rm CO}$ ratio.
We would pay attention to the negative-velocity component of the EMR,
which is prominently seen in $\ell$-$v$ diagrams
at $b=0\fdg 0, -0\fdg 2, -0\fdg 4$ (Fig.\ \ref{fig-lvall}).
Higher ratio is more extensively seen in the positive longitudes.
For example, we present the longitudinal distribution of the CO emission
and the OH absorption at $b=-0\fdg 2$ in Figure \ref{fig-emr}.
The Figure shows the trend that the OH absorption is deeper
in the positive longitudes compared with the CO emission.
This suggests that the EMR, at least its negative-velocity component,
is inclined so that the positive-longitude end is closer to us.
This geometry agrees with the interpretation by \citet{binney1991}
rather than the axisymmetric `expanding ring'.

\subsection{Physical conditions}
\citet{sawada2001} have investigated $\ell$-$v$ distribution of
physical conditions of
molecular gas in the Galactic centre using the CO intensity ratios
${}^{12}{\rm CO}\; J=2-1/{}^{12}{\rm CO}\; J=1-0$
[$R_{2-1/1-0}(^{12}{\rm CO})$] and
${}^{13}{\rm CO}\; J=2-1/{}^{12}{\rm CO}\; J=2-1$
[$R_{13/12}(J=2-1)$].
In this subsection we discuss the face-on distribution of CO intensity
ratios and physical conditions of the gas derived from the ratios.

In the face-on map derived in {\S} \ref{subsec-result}
(Fig.\ \ref{fig-result}), the Sgr B cloud complex at
$(\ell,v_{\rm LSR}) \simeq (0\fdg 6, 50\,{\rm km\, s^{-1}})$ is
mapped on to $(x,y) \simeq (90\,{\rm pc},-100\,{\rm pc})$.
That is highly offset from the other part of the central condensation toward
us, and the isovelocity contours are unnaturally curved around there.
This is caused by the intense discrete continuum sources Sgr B1 and B2.
Therefore we reassigned the cloud positions toward $\ell = 0\fdg 6$
by interpolating the velocity field from $\ell = 0\fdg 4$ and $0\fdg 8$:
i.e., data at each velocity bin toward $\ell = 0\fdg 6$ are placed  on to the
position of the corresponding velocity in the interpolated velocity field.
The resultant face-on maps of CO and corresponding radial velocity
are presented in Figures \ref{fig-intratio}{\it a\/} and
\ref{fig-intratio}{\it b\/}, respectively.
\begin{figure*}
\includegraphics[width=126mm]{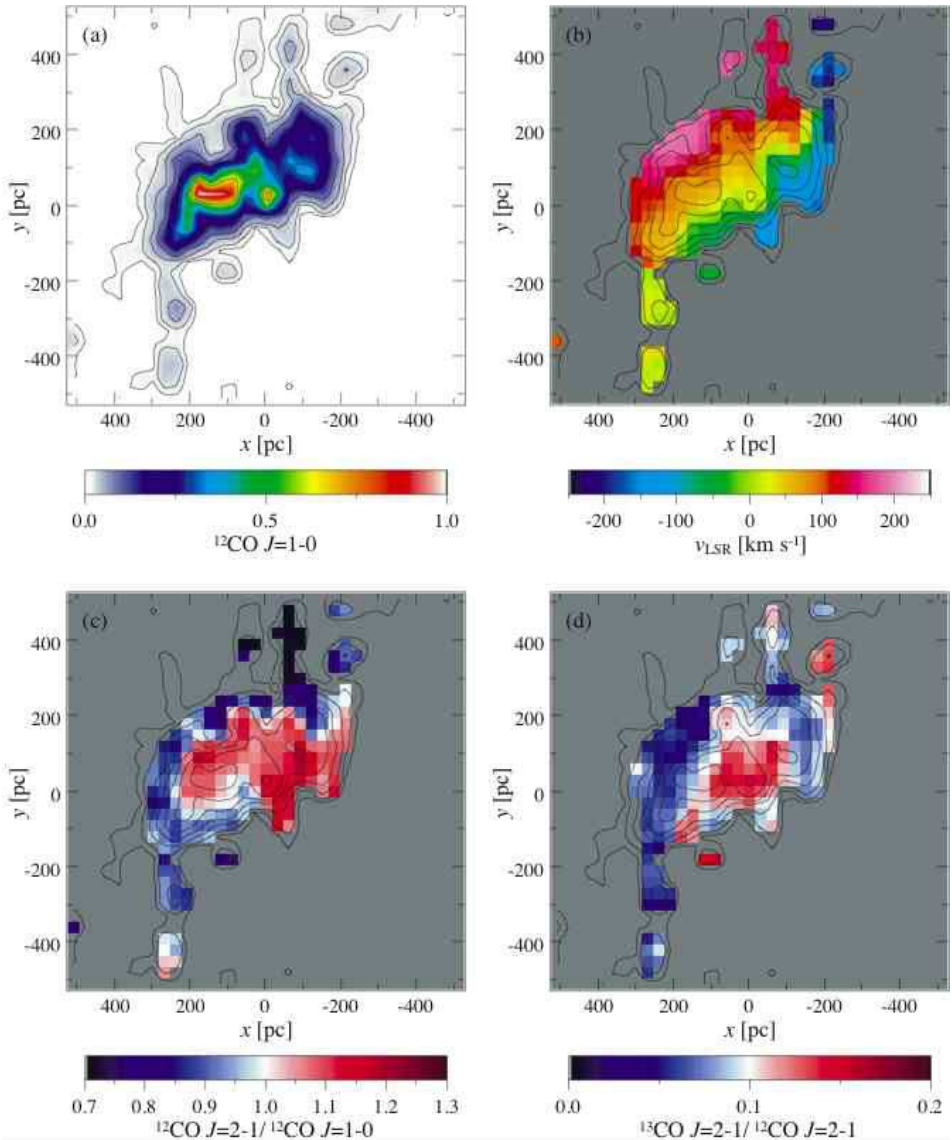}
\caption{Face-on distribution (at $b=0\fdg 0$) of 
({\it a\/}) ${}^{12}{\rm CO}\; J=1-0$;
({\it b\/}) Radial velocity;
({\it c\/}) ${}^{12}{\rm CO}\; J=2-1/{}^{12}{\rm CO}\; J=1-0$ intensity
ratio; and
({\it d\/}) ${}^{13}{\rm CO}\; J=2-1/{}^{12}{\rm CO}\; J=2-1$ intensity
ratio.
Values are shown in the region where the ${}^{12}{\rm CO}\; J=1-0$
emission is more intense than 2 per cent of the peak.
Contours show the ${}^{12}{\rm CO}\; J=1-0$ distribution:
levels are 0.01, 0.02, 0.04, 0.08, 0.16, 0.32, and 0.64 of the peak.}
\label{fig-intratio}
\end{figure*}

The velocity field in Fig.\ \ref{fig-intratio}{\it b\/} can be used to
convert an $\ell$-$v$ plot of any parameter (intensity of other lines,
line intensity ratios, etc.) to a face-on view.
We projected the ${}^{12}{\rm CO}\; J=2-1$ and ${}^{13}{\rm CO}\; J=2-1$
data \citep{sawada2001} at $b=0\fdg 0$ on to the $x$-$y$ plane.
Figures \ref{fig-intratio}{\it c\/} and \ref{fig-intratio}{\it d\/} show the
face-on distribution of intensity ratios
$R_{2-1/1-0}(^{12}{\rm CO})$ and
$R_{13/12}(J=2-1)$, respectively.
On one hand, $R_{2-1/1-0}(^{12}{\rm CO})$ is high (1.0--1.2)
almost all over the central condensation.
On the other hand, high $R_{13/12}(J=2-1)$ (i.e., 0.10--0.15)
is seen mainly within a radius of $\simeq 100\,{\rm pc}$ around the centre.

Using one-zone large velocity gradient
\citep[LVG; see, e.g.,][]{goldreich1974}
analysis, \citet{sawada2001} derived the physical conditions of the gas from
these intensity ratios.
Though the derived physical conditions depend on the assumed gas kinetic
temperature $T_{\rm k}$, the thermal pressure $p/k = n({\rm H_2})T_{\rm k}$
[$n({\rm H_2})$ is the derived number density of molecular hydrogen] was
found to be almost independent of the assumed $T_{\rm k}$.

\begin{figure*}
\includegraphics[width=126mm]{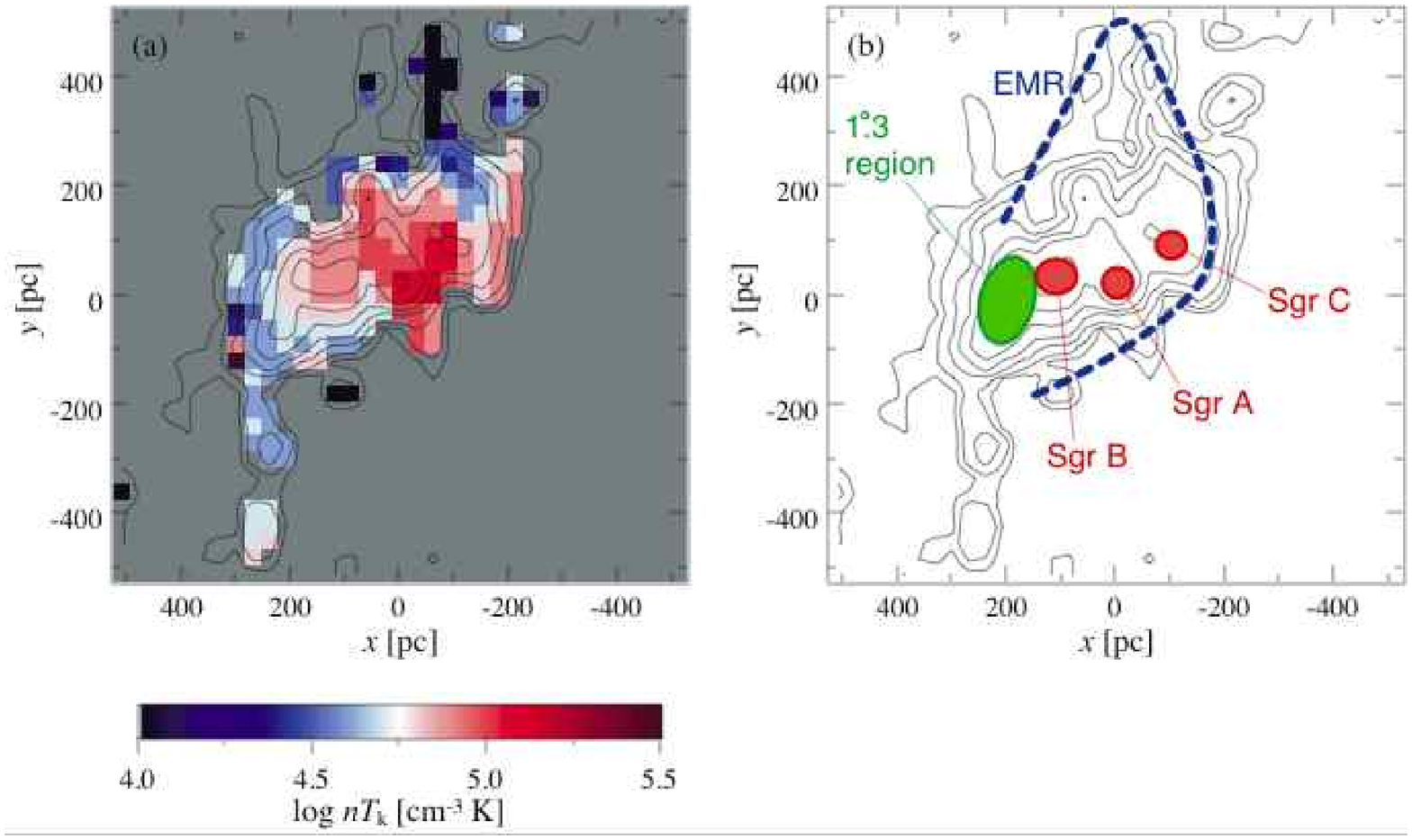}
\caption{({\it a\/}) Face-on distribution of gas thermal pressure derived
from one-zone LVG analysis
at $b=0\fdg 0$.
Values are shown in the region where the ${}^{12}{\rm CO}\; J=1-0$
emission is more intense than 2 per cent of the peak.
({\it b\/}) Schematic illustration of some molecular features.
Contours in both panels are the same as those in Fig.\ \ref{fig-intratio}.}
\label{fig-physcon}
\end{figure*}
Figure \ref{fig-physcon}{\it a\/} shows the distribution of
pressure $n({\rm H_2})T_{\rm k}$ obtained in the same way. 
The highest pressure, $10^{4.9}$--$10^{5.1} \,{\rm cm^{-3}\, K}$, is seen
within the central $100\,{\rm pc}$.
It is noteworthy that the highest pressure is found roughly symmetrically
around the centre, whereas the molecular mass distribution is
asymmetric in the way that the larger fraction of the CO emission comes from
the positive longitude (see Fig.\ \ref{fig-intratio}{\it a\/}).
The central region corresponds to the `high pressure region' in the
$\ell$-$v$ diagram described in \citet{sawada2001}.
\citet*{launhardt2002} investigated the inner hundreds of parsecs of the Galaxy
using {\it IRAS} and {\it COBE} data, and deduced that gas and dust within
the galactocentric radius of $120\,{\rm pc}$ (referred to as
`inner warm disk' in their paper) are warmer than those at larger radii
(`outer cold torus').
The `high pressure region' found in our analysis may correspond to the
`inner warm disk' in \citet{launhardt2002}.
The $1\fdg 3$ region shows slightly lower pressure than the average of
the central condensation; and the EMR has even lower pressure.

\begin{figure}
\includegraphics[width=63mm]{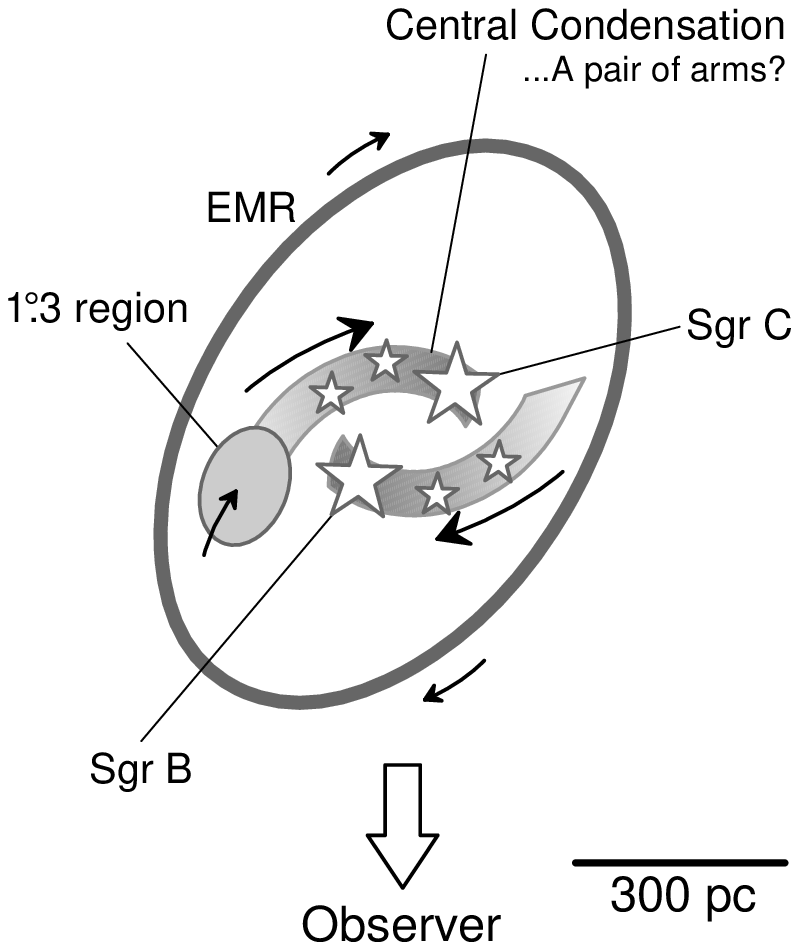}
\caption{A picture of molecular gas in the Galactic centre region
proposed on the basis of the results from this work.
Linear scales are approximate.}
\label{fig-sumpunch}
\end{figure}
A schematic illustration of the major molecular features is shown
in Figure \ref{fig-physcon}{\it b}.
\citet{sofue1995} identified a pair of arm-like features in
the central condensation from the ${}^{13}{\rm CO}$ data taken by
\citet{bally1987},
though these `arms' are not clearly separated in our face-on map.
It may be due to insufficient spatial resolution of the present analysis
and to deviation from the assumed smooth, axisymmetric
continuum emissivity because of embedded discrete radio continuum sources.
These arms and the high pressure region occupy a similar location
in the $\ell$-$v$ diagram \citep{sawada2001}.
\citet{oka1996} argued that formation of massive stars may have been taking
place continuously or intermittently in this region for more than $10^8$
years based upon comparisons between their CO data and the $\ell$-$v$
distributions of ${\rm H}109\alpha$ recombination line emission 
\citep{pauls1975}
and OH/IR stars \citep{lindqvist1991},
and called it the `star-forming ring'.
We consider that concentration of gas may occur to form high pressure
(high density and/or high temperature) clouds in this region,
from which stars form.
\citet{mizutani1994} observed the [C\,{\sc ii}] $158\,\umu{\rm m}$
line, which is considered to trace photodissociation regions,
within $|\ell| \le 0\fdg 7$ and found that the emission has peaks
at Sgr B1 and Sgr C, in addition to Sgr A region.
This might suggest that the star-forming activity is enhanced in the
Sgr B and C regions.
In the two-arm model by \citet{sofue1995},
Sgr B and C are both located on the leads of these arms.
In his analysis, `Arm I' (which contains Sgr B) with lower
(rather approaching) radial velocity is in the near side;
`Arm II' with Sgr C is in the far side.
In our face-on map, the far side of the central condensation shows
larger receding velocity (see {\S} \ref{subsec-result}).
This geometry is consistent with that suggested by \citet{sofue1995},
therefore Sgr B and C might be actually at the leads of these arms.
The geometry that the sites of active star formation are located at the
leads of inner arms may suggest a time lag between the arrival of the gas
into the central orbit and the beginning of star formation as discussed by
\citet*{kohno1999} for NGC 6951.

The 1\fdg 3 region lies at the Galactic-eastern end of
the central condensation.
\citet{oka1998b} noted that the large velocity extents and the
fluffy structure of the clouds and filaments in this region
may be a sign that these clouds are formed by large-scale shocks.
\citet{huttemeister1998} reported that SiO in this region
is exceedingly abundant and its emission arises from hot, thin gas;
which also indicates the existence of shocks.
From their arguments and its geometry,
we suppose that the gas in the $1\fdg 3$ region has recently arrived in
the central region and collided with the central condensation.
The extremely large scale height of this region
compared with other parts of the central condensation
supports our idea: the gas which has just fallen into the
central gravitational field has not relaxed yet.
The envelope of the $1\fdg 3$ region toward the Galactic-east and
smaller radial velocity is placed closer to us than the main body of the
$1\fdg 3$ region in the face-on map.
This may support our idea that the $1\fdg 3$ region has just bumped
into the central condensation: the envelope may be a tail of the cloud
that has not yet reached the condensation.
\citet*{sakamoto2000} suggested that `episodic fueling' is occurring
in NGC 5005: the $1\fdg 3$ region might be a result of such kind of
cloud infall.
Our results are schematically summarized in Figure \ref{fig-sumpunch}. 

We have found that the major axis of the central condensation is inclined
by an angle of $\simeq 70\degr$ with respect to the line of sight.
Therefore, the position angle between the major axes of the large-scale
stellar bar and the central condensation is $\simeq 40\degr$--$50\degr$
(central condensation is leading the bar), if our
viewing angle of the bar is $\simeq 20\degr$--$30\degr$
\citep[see][and references therein]{morris1996,gerhard1999,deguchi2002}.
In some barred galaxies, there are central molecular gas concentrations
\citep[`twin peaks',][]{kenney1992}.
In a face-on projection of high resolution images,
some of these concentrations tend to consist of a pair of arms and
their major axes are inclined with respect to the bar major axes by
moderate angles:
such as IC 342 by \citet{ishizuki1990},
M101 etc.\ by \citet{kenney1992},
NGC 1530 by \citet{reynaud1999},
NGC 6951 by \citet{kohno1999},
NGC 5383 by \citet{sheth2000}, and
NGC 4303 by \citet{schinnerer2002}.
They are similar to our face-on view of the Galactic centre.
Hydrodynamical simulations of the barred galaxies also produce
similar gas distributions: e.g., simulations for NGC 1365
made by \citet{lindblad1996} produce a central condensation
which is inclined from the bar axis with a similar angle as that
we have obtained, $40\degr$--$50\degr$.
\citet{binney1991} interpreted the longitude-velocity feature
that corresponds to the central condensation as a projection of
gaseous materials in $x_2$ orbits.
However the $x_2$ orbits are aligned perpendicular to the
potential \citep[Fig.\ 3 in][]{binney1991}.
As a result, the sense of inclination with respect to the line of sight
are opposite between our and their interpretation.
This difference may be due to the fact that the $x_2$ orbits
perpendicular to the bar are stellar (collisionless) orbits.
\citet{wada1994} showed from his damped-orbit model that gaseous orbits
do not align with respect to the bar but their major axes gradually
changes with the Galactocentric radius.
He showed that the orbits at the Inner Lindblad Resonances
lead the bar by $45\degr$: the trend agrees with the gas distribution
we have obtained.

\subsection{Validity of parameters}\label{subsec-paramvalid}
\subsubsection{Beam filling factor}\label{subsubsec-validf}
We have assumed that $f$, the beam filling factor of the OH absorbing gas,
is equal to unity.
From Eq.\ (\ref{eq-tauapp}), we can show $f > \tau_{\rm app}$:
$f$ gets the lowest if the cloud is located in the solar neighborhood
and has infinite opacity.
For some lines of sight, the apparent opacity $\tau_{\rm app}$,
which is the lower limit of $f$, is rather large:
0.64 toward Sgr B2; 0.35 toward the $1\fdg 3$ region and Clump 2.
This fact supports that the beam filling factor is large and can be
reasonably replaced with unity.

\citet{sawada2001} estimated the
beam filling factor of ${\rm CO}\; J=1-0$ emission to be $0.4$--$0.7$
based upon the LVG analyses and high resolution CO data taken by
\citet{oka1998b}.
Moreover, since it is expected that the OH absorption arises also from
less dense cloud envelope compared with the CO emission,
the beam filling factor of OH absorbing gas is at least similar and can be
even larger.
This again suggests that $f$ is nearly unity.

\subsubsection{Excitation temperature}
We have assumed that the excitation temperature of OH,
$T_{\rm ex}({\rm OH})$, is uniformly $4\,{\rm K}$ above the CMB.
It is reported that $T_{\rm ex}({\rm OH})$ typically ranges in
$5$--$10\,{\rm K}$ \citep{elitzur1992}.
Our value, $4\,{\rm K} + T_{\rm CMB}$, is close to this.

\begin{figure}
\includegraphics[width=52mm]{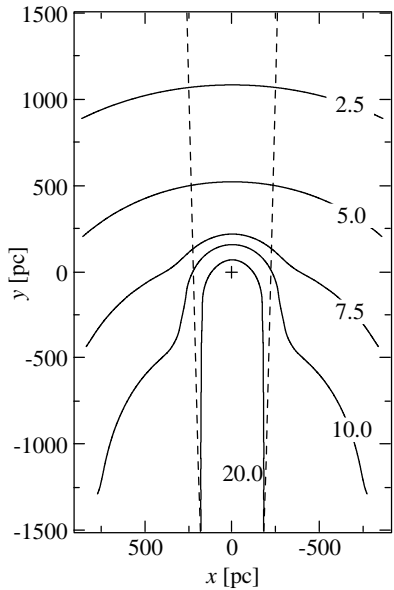}
\caption{Contours of integrated continuum emissivity
$\int_{-\infty}^{s_0} j(r) {\rm d}s$ (see Eq.\ (\ref{eq-tauapp}))
at $b=0\fdg 0$ based upon the 3 Gaussians model.
Contour levels are 2.5, 5.0, 7.5, 10.0, and 20.0 K in units of
brightness temperature.
A molecular cloud located on this plane should be seen in emission
if $T_{\rm ex}({\rm OH})$ is higher than the integrated continuum
emissivity shown by the contours. 
Coordinates are the same as Fig.\ \ref{fig-result}.
The centre $(x,y)=(0,0)$ is shown by a cross.
Dashed lines indicate $\ell = \pm 1\fdg 5$.}
\label{fig-contcont}
\end{figure}
Figure \ref{fig-contcont} shows contours of integrated continuum emissivity
along lines of sight, $\int_{-\infty}^{s_0}j(r) {\rm d}s$ in Eq.\ (\ref{eq-tauapp}),
at $b=0\fdg 0$.
It is expected that the gas behind the contour
which is equal to $T_{\rm ex}({\rm OH})$ is seen in emission,
while the gas in front of the contour is seen in absorption.
However, no significant OH emission is observed in this region.
Thus, if $T_{\rm ex}({\rm OH}) \ga 6\,[{\rm K}]$,
the face-on gas distribution is almost restricted to near side of the
Galactic centre, $y<0$.
We consider it is unrealistic.
On the other hand, \citet{boyce1994} noted that the OH absorption was
only detected at positions where the continuum temperature exceeded
$5\,{\rm K}$ of antenna temperature ($7.5\,{\rm K}$ of brightness
temperature).
Lower limit of excitation temperature of OH is provided from this fact:
$T_{\rm ex}({\rm OH})$ is higher than the integrated continuum emissivity
if the OH lines are not seen as absorption.
For a cloud in the Galactic centre region, integrated continuum emissivity
at the position of the cloud is typically a half of the observed brightness
temperature.
Thus $T_{\rm ex}({\rm OH})$ should not be far below $(7.5/2)\,{\rm K}$.

Again using Eq.\ (\ref{eq-tauapp}),
we can show that the upper limit of $T_{\rm ex}({\rm OH})$ is described as
$(1-\tau_{\rm app})\int_{-\infty}^{s_\odot} j(r){\rm d}s$:
$T_{\rm ex}({\rm OH})$ gets the highest if
the cloud is located in the solar neighborhood.
Toward Clump 2, this value goes down to $7\,{\rm K}$,
which gives a similar constraint to
$T_{\rm ex}({\rm OH}) \la 6\,[{\rm K}]$ derived above.
On the other hand, our trials have shown that 
almost all clouds are placed on the far side of the Galactic centre
($y>0$) at high Galactic latitudes
if $T_{\rm ex}({\rm OH})$ is lower ($\la 3\,[{\rm K}]$).

In conclusion, $4\,{\rm K}$ is a proper estimate of
the excitation temperature.

\subsubsection{The `$Z$' factor}
We have derived $Z$, the factor to convert $T_{\rm CO}$ into
$\tau_{\rm OH}$, in {\S} \ref{sec-derivedist} empirically.
Here we attempt an independent estimate of $Z$.

There are three previously-known relations.
(1)
It is considered that the column density of
molecular hydrogen $N({\rm H_2})$
is proportional to the ${\rm CO}\; J=1-0$ integrated intensity.
The conversion factor (the so-called $X$-factor) is measured to be about
$2\times 10^{20}\,[{\rm cm^{-2}\, (K\, km\, s^{-1})^{-1}}]$ for
molecular clouds in the Galactic disc \citep*[see, e.g.,][]{dame2001}.
For the Galactic centre clouds, however, it is reported that the $X$-factor
is smaller than that for the Galactic disc:
i.e., $X = {\rm several} \times 10^{19}$
\citep[see, e.g.,][]{oka1998a}.
(2)
Relative abundance of OH to molecular hydrogen,
$[{\rm OH}]/[{\rm H_2}]$, is typically ${\rm several}\times 10^{-7}$
\citep[see, e.g.,][]{herbst1989}.
(3)
The OH column density is derived from the OH $1667\,{\rm MHz}$
opacity $\tau$ as
$N({\rm OH}) = 2\times 10^{14} T_{\rm ex} \int\tau {\rm d}v$.

Using these relations, the $Z$ factor is written as
\begin{eqnarray}
Z\,[{\rm K^{-1}}] =
\frac{[{\rm OH}]/[{\rm H_2}]}{2\times 10^{14}}
\frac{X\,[{\rm cm^{-2}\, (K\, km\, s^{-1})^{-1}}]}{T_{\rm ex}\,[{\rm K}]} \\
= 5\times 10^{-3}
\frac{[{\rm OH}]/[{\rm H_2}]}{10^{-7}}
\frac{X}{10^{20}\,{\rm cm^{-2}\, (K\, km\, s^{-1})^{-1}}}
\frac{10\,{\rm K}}{T_{\rm ex}} .
\end{eqnarray}
This value becomes ${\rm several}\times 10^{-2}$, which is several times
smaller than that we have adopted.
However $Z \simeq {\rm several}\times 10^{-2}$ does not seem to be
realistic, because such a small $Z$ cannot reproduce the observed
absorption depth for a significant fraction of the clouds.
This discrepancy can be solved if we consider larger $X$
(${\rm several}\times 10^{20}$) or higher OH abundance
(${\rm several}\times 10^{-6}$).
On one hand, the large $X$-factor is unrealistic because the small
$X$-factor in the Galactic centre region is consistently suggested in
various ways \citep[see, e.g.,][]{blitz1985,cox1989,sodroski1995,oka1998a}.
On the other hand, it is reported that the OH abundance is
enhanced behind interstellar shocks \citep[see, e.g.,][]{wardle1999}.
In a high resolution CO map of the Galactic centre,
innumerable shells/arcs are seen \citep{oka1998b,oka2001a,oka2001b}.
If the shells/arcs are formed by a number of explosive phenomena such as
supernova, the Galactic centre is filled with shocks.
In such condition, the OH abundance may be higher than that in
usual condition.
High ionization rate caused by X-ray also enhances the OH abundance
\citep{lepp1996}.

\subsubsection{One-to-one correspondence between position and velocity}
In {\S} \ref{sec-derivedist}, we have assumed that a value in the radial
velocity corresponds to a single position at a given line of sight.
Of course this assumption is nominally wrong: there are two (or more)
points in a line of sight with a given radial velocity.
Thus, if the Galactic centre region is almost filled with molecular gas, 
the assumption becomes fairly wrong.
Molecular gas distribution is, however, rather localized.
In fact, our (Fig.\ \ref{fig-ohcoratio}) and higher-resolution
\citep[see, e.g.,][]{oka1998b} molecular line data show that velocity
overlapping between major features (Sofue's arms, EMR, etc.) occurs in
very restricted regions.
Therefore the assumption might not introduce serious errors into
the resultant face-on maps.

Even if overlapping of clouds is occurring, the results do not change
very much:
let we assume that two individual clouds (positions $s_1$ and $s_2$;
$s_1 < s_2$) with the same velocity is seen as a single cloud.
Position of this `cloud' is derived as $s_0$ in our method.
Using Eq.\ (\ref{eq-tauapp}), we can show $s_1 < s_0 < s_2$
if the ${\rm H_2}$ column density can be fully traced by $T_{\rm CO}$.
Thus the overall CO distribution derived is not heavily affected
in this situation.

\section{CONCLUSIONS}
We have developed a method to derive positions of molecular clouds
along the lines of sight.
The method is completely independent of any kinematic model and
based on observable data alone;
the CO emission line, the OH absorption line,
and $18\,{\rm cm}$ continuum distribution.
It is applied to the central region of the Milky Way
to obtain a molecular face-on map.
The obtained spatial and kinematical structure of molecular gas
shows the following characteristics.

\begin{itemize}
\def\labelitemi{\bf --}
\item The CO distribution mainly consists of a central condensation and
a ridge.
\item The central condensation is elongated and its major axis is
inclined with respect to the line of sight by $\simeq 70\degr$ so that
the Galactic-eastern end is closer to us.
The gas within it shows
highly noncircular motion: the gas in the far side is receding whereas
the gas in the near side is approaching.
This noncircularity of the gas motion is most likely induced by a barred
potential.
\item The ridge, corresponding the Bania's Clump 2, lies closer to
us than the central condensation by $\simeq 1\,{\rm kpc}$.
This feature apparently mimics offset ridges often seen in barred galaxies,
although its stretch along our line of sight might be artificial.
\item The so-called `Expanding Molecular Ring' feature does not
appear as a coherent structure on the resultant face-on map.
However its negative-velocity side, which lies in the near side of
the Galactic centre, is inclined so that the Galactic-eastern side
is closer to us.
This is consistent with the arguments by \citet{binney1991}, that 
the EMR is a projection of $x_1$ orbit in a barred potential,
rather than the original picture of a ring that is expanding and rotating.
\end{itemize}

These results give a new evidence for
the existence of a bar in the Milky Way Galaxy
based on direct distance derivation independent of kinematic models.
A face-on map of thermal pressure derived from comparison of
three CO transition lines and an LVG analysis is also presented.
The pressure is distinctly higher (i.e., $\ga 10^5\,{\rm cm^{-3}\, K}$)
within the Galacto\-centric radius of $100\,{\rm pc}$,
compared with the outer region.
This high pressure region coexists with `Galactic centre arms'
\citep{sofue1995} and `star-forming ring' \citep{oka1996}.
Concentration of clouds would be occurring due to some kind of gas orbit
crowding.

\section*{Acknowledgments}

We acknowledge Leonardo Bronfman for providing us
the ${\rm CO}\; J=1-0$ data in a computer readable form.
This work was supported by a Grant-in-Aid for Scientific Research
of the Ministry of Education, Culture, Sports, Science, and Technology
08404009 and 10147202.

\label{lastpage}

\end{document}